\documentstyle[12pt,titlepage]{article}

\advance\textwidth1cm
\advance\textheight2.5cm
\advance\oddsidemargin-0.5cm
\advance\topmargin-1cm

\newcommand{\be}{\begin{equation}}
\newcommand{\ee}{\end{equation}}
\newcommand{\bea}{\begin{eqnarray}}
\newcommand{\eea}{\end{eqnarray}}
\newcommand{\q}{{\bf q}}
\newcommand{\p}{{\bf p}}
\newcommand{\Q}{{\bf Q}}
\renewcommand{\P}{{\bf P}}
\newcommand{\M}{{\bf M}}
\newcommand{\R}{{\bf R}}
\newcommand{\RP}{\dot{{\bf R}}}
\newcommand{\C}{\tilde{C}}
\renewcommand{\H}{{\bf H}}
\newcommand{\HP}{\dot{{\bf H}}}
\newcommand{\pa}{\partial}
\newcommand{\KQ}{\vec{K}^2}
\newcommand{\KQS}{\vec{{\cal K}}^2}
\newcommand{\JQ}{\vec{J}^2}
\newcommand{\IQ}{\vec{I}^2}
\newcommand{\NQ}{\vec{N}^2}
\newcommand{\Yjmml}{Y_{jm_1m_2}^l}
\newcommand{\Yf}{Y_{j_1m_{11}m_{12}j_2m_{21}m_{22}}^f}
\newcommand{\YK}{Y_{j_1m_1j_2m_2KM}^l}
\newcommand{\ir}{irreducible}
\newcommand{\rep}{representation}
\newcommand{\Hp}{{\cal H}_{phys}}

\renewcommand{\S}{{\cal S}}
\newcommand{\OS}{\tilde{{\cal O}}}
\newcommand{\QOS}{{\cal Q}\tilde{{\cal O}}}
\newcommand{\QS}{{\cal QS}}

\begin{document}

\begin{titlepage}

\vspace*{-2cm}
\begin{flushright} { THEP 95/17\\Universit\"at Freiburg\\August 1995\\
                     hep-th/9510019}
\end{flushright}
\vspace{0.5cm}
\begin{center}
{\LARGE\bf The five-dimensional Kepler Problem \\[7mm]
           as an SU(2) Gauge System: \\[7mm]
           Algebraic Constraint Quantization}\\[2.5cm]
{\large Michael Trunk}\\[4mm]
{Universit\"at Freiburg\\
Fakult\"at f\"ur Physik\\
Hermann--Herder--Str.\ 3\\
D--79104 Freiburg\\
Germany\\[5mm]
e-mail: trunk@phyv3.physik.uni-freiburg.de}\\[3.5cm]
{\bf Abstract}\\[5mm]
\end{center}

Starting from the structural similarity between the quantum theory of gauge
systems and that of the Kepler problem, an SU(2) gauge description of the
five-dimensional Kepler problem is given. This non-abelian gauge system is
used as a testing ground for the application of an algebraic constraint
quantization scheme which can be formulated entirely in terms of observable
quantities. For the quantum mechanical reduction only the quadratic Casimir
of the constraint algebra, interpreted as an observable, is needed.
\end{titlepage}

\section{Introduction}

In this paper we present an example for the quantization and
reduction of a non-abelian gauge system along the guidelines of
an algebraic constraint quantization scheme which can be formulated
entirely in terms of observable quantities. This algebraic scheme does
not make use of the individual constraints as projectors onto the physical
subspace of an extended Hilbert space. Rather, it treats the intrinsically
defined observable content of the constraint algebra, comprised in its
Casimir elements, as \rep\ conditions to determine the physical \rep s
of the algebra of observables. A heuristic formulation of the procedure,
starting from a classical first class constrained system, is as follows.
\begin{itemize}
\item Define the algebra of observables $\cal O$ as the strong Poisson
commutant
of the constraint algebra $\cal C$. By definition of $\cal O$, the Casimirs
of $\cal C$ must be addressed as observables. At the same time they are
also Casimirs of $\cal O$ and especially of the observable symmetry or
invariance algebra $\cal S$ of the system.
\item In general there exist functional relations between the Casimirs of
$\cal C$ and those of $\cal O$ ($\cal S$), which allow to express the
Casimirs of $\cal C$ as functions of the generators of $\cal O$ ($\cal S$).
\item As a first step towards the quantization of the system, choose
two subalgebras of $\cal O$: $i$) the symmetry algebra $\cal S$, which
contains a good deal of the structural information about the system, and,
$ii$) an algebra $\OS$ which should allow to generate $\cal O$ and to
reproduce the dynamical content of the system. Of course, (part of) $\S$
can be contained in $\OS$ to form a dynamical or non-invariance algebra
for the system. $\S$ as well as $\OS$ are characterized by the Poisson
commutation relations of and the functional dependencies between their
generators.
\item The next step consists in constructing the quantum analogs $\QS$
and $\QOS$ of $\S$ and $\OS$, i.e.\ the commutator algebras which
correspond to them. As the connection between the classical and quantum
theories is closest on the algebraic level, we shall require that $\QS$,
as a commutator algebra, be isomorphic to $\S$. As to the relationship
between $\OS$ and $\QOS$, we cannot a priori require them to be isomorphic.
This is possible only in special cases, depending on the physical
interpretation: For example if $\OS$ can be chosen to consist of globally
defined canonical coordinates on the reduced phase space or if it forms
a dynamical Lie algebra. Nevertheless, the algebraic structure of $\OS$
cannot be drastically changed, and that of $\QOS$ should be as close as
possible to that of $\OS$. In any case the covariant transformation
properties of its generators with respect to $\S$ should be preserved.
Also, for the theory to possess the correct classical limit, in leading order
in $\hbar$ (as far as explicit powers of $\hbar$ are concerned, cf.\ the next
item) the commutator algebra must be isomorphic to the Poisson bracket algebra.
\item The crucial step is the determination of the functional relations
between the Casimirs of $\cal C$, $\QS$ and $\QOS$, and thereby of the
expressions for the Casimirs of $\cal C$ in terms of the generators of
$\QS$ or $\QOS$. As the generators of $\QS$ and $\QOS$, and therefore their
Casimirs, are to carry a definite gradation in terms of physical dimensions,
possible correction terms, which must be formed from Casimirs and carry
explicit powers of $\hbar$, can be largely restricted by a dimensional
analysis. In leading (i.e.\ zeroth) order in (explicit powers of) $\hbar$
the classical expressions must be reproduced.
\item The final step is the construction of the \ir\ \rep s of $\QS$ and
$\QOS$ on a suitably chosen Hilbert space $\cal H$. The system being
constrained means that $\cal H$ cannot be irreducible with respect to $\QOS$
and that it contains unphysical \rep s of $\QOS$ and $\QS$. The ``vanishing"
of the constraints has two consequences. It implies the projection onto
the zero--eigenspaces of the Casimirs of $\cal C$, and, via the functional
dependencies between the Casimirs of $\cal C$ and those of $\QOS$ and
$\QS$, it induces relations which must be satisfied by the Casimirs of
$\QOS$ or $\QS$ respectively.
\item The selection of the physical \rep s of $\QOS$ and $\QS$ is
achieved by means of the eigenvalues of and the relations between their
Casimir operators. That is, the physical Hilbert space is spanned
by those \ir\ \rep s of $\QOS$ and $\QS$, in which the Casimirs of
$\cal C$ vanish and the induced relations are satisfied. In many cases,
to determine the physical Hilbert space, it will be sufficient and more
favourable to restrict the procedure to the symmetry algebra alone,
because its \rep s are often well studied and easier to handle.
\end{itemize}
Note that the application of the algebraic scheme does not require the
quantization of the unphysical constraint algebra (although it may
facilitate the analysis if we can consistently quantize it).

If the above reasoning is correct, it should be a common feature
of the quantum theory of gauge systems, that their physical Hilbert
space carries only a restricted class of the \ir\ \rep s of the symmetry
algebra of the system, which is determined by the characteristic
identities satisfied by the Casimirs of the symmetry algebra. This
property is shared by the quantum theory of the Kepler problem (KP)
in any number of dimensions $\ge 3$. This structural similarity suggests
that it should be possible
to describe the KP as the reduced form of a higher-dimensional gauge system.
In ref.\ \cite{Tr 94} this has been done successfully for the
three-dimensional KP with SO(2) as the gauge group. There it has been
shown that the application of the algebraic constraint quantization
scheme to the corresponding gauge system yields the well known quantum
theory of the hydrogen atom. Of course, for an abelian constraint algebra
the Casimirs are the constraints themselves, and there are as many
Casimirs as there are constraints. But in the case of a non-abelian constraint
algebra the number of Casimirs is less than the number of constraints,
and it has to be shown that the information contained in the Casimirs
is sufficient for the reduction of the system.

In the present paper we will give a description of the five-dimensional
Kepler problem (KP$_5$) as an SU(2) gauge system by application of the
so-called Hurwitz transformation. It will be shown that the reduction of the
resulting gauge system according to the algebraic constraint quantization
scheme, which uses only the quadratic Casimir of the constraint algebra
su(2), yields the quantum theory of the KP$_5$.

The organization of the paper is as follows. In section 2 we briefly
describe the symmetry algebra of the classical KP$_5$ and the characteristic
relations obeyed by its Casimirs. In section 3 the Hurwitz transformation
is introduced and applied to the KP$_5$ in section 4. In section 5 we present
the quantum theory of the KP$_5$ and determine the \rep s of its symmetry
algebra. The quantization and reduction of the corresponding  SU(2) gauge
system is carried out in section 6, and the resulting quantum theory is
compared to the quantum theory of the KP$_5$. The appendix contains a brief
characterization of the unitary \ir\ \rep s of the groups SO(6), E(5) and
SO(5,1).

\newpage
\section{The symmetry algebra of the five-dimensional Kepler problem}

The five-dimensional Kepler problem (KP$_5$) is the dynamical system
($P,\omega,H$), with phase space $P=T^*\RP^5=\R^5\times\RP^5$
($\RP^5=\R^5\setminus \bigl\{  0 \bigr\} $), symplectic form $\omega =d\q
\wedge d\p$, and Hamiltonian
\be H=\frac12\: \p^2 - \,\frac{k}{q} ,\quad\qquad k>0, \qquad q=\sqrt{\q^2}.\ee
The constants of motion are the components of the angular momentum tensor
\be L_{ij}=q_i p_j - q_j p_i ,\qquad\qquad 1\le i,j \le 5 \ee
and of the Lenz--Runge vector
\be \M = (\frac12\: \p^2 + H)\,\q - (\q\cdot\p)\,\p. \ee
The Poisson commutation relations of $H$, $L_{ij}$ and $M_i$
\be \big\{H,L_{ij}\big\} = 0 = \big\{H,M_i\big\} \qquad\;
\big\{L_{ij},L_{ik}\big\} = L_{jk} \label{sa1}\ee
\be \big\{L_{ij},M_i\big\}=M_j \qquad\qquad \big\{M_i,M_j\big\} = -2HL_{ij}
\label{sa2}\ee
define an abstract algebra $\cal S$. The algebra $\cal S$ possesses
three algebraically independent Casimir invariants, which can be obtained
in the following way. Let
\be l_{ij} = \sqrt{-2H}\: L_{ij},\qquad\quad l_{i6}=-l_{6i}=M_i,
\ee
then, for constant $H\neq 0$, $l_{\mu\nu}$, $1\le\mu,\nu\le 6$, generate a
deformation of the Lie algebra so(6), which is characterized by the
commutation relations
\be \big\{l_{\mu\nu},l_{\mu\rho}\big\}=\sqrt{-2H}\: l_{\nu\rho}.
\ee
Irrespective of the sign of the square root, the Casimir invariants of
this algebra are also Casimirs of the algebra $\cal S$, even if $H$ is
not kept constant. They are built in the same way as the Casimir invariants
of so(6) and can be chosen as
\bea \C_2 &=& \frac12\: l_{\mu\nu} l_{\mu\nu}=\M^2 - 2HL^2 \qquad\qquad\qquad
              \quad (L^2=\frac12\: L_{ij}L_{ij})\\
     \C_3 &=& \frac1{-2H}\: \varepsilon_{\mu\nu\rho\sigma\tau\upsilon}
              l_{\mu\nu}l_{\rho\sigma}l_{\tau\upsilon}=6\,\varepsilon_{ijklm}
              L_{ij}L_{kl}M_m\\
     \C_4 &=& \frac12\: l_{\mu\nu}l_{\nu\rho}l_{\rho\sigma}l_{\sigma\mu}=
              2H^2 L_{ij}L_{jk}L_{kl}L_{li} + 4HM_iM_jL_{jk}L_{ki} + (\M^2)^2.
\eea
For $H=0$ the right hand sides of these expressions are still well defined,
but $\C_4$ becomes algebraically dependent on $\C_2$, and we will
use instead
\bea \C_4^{(0)} &=& \M^2 L^2 - \frac14\, (M_i L_{ij}+L_{ij} M_i)(M_k L_{kj}+
                    L_{kj} M_k)= \nonumber \\
                &=& \M^2 L^2 + M_i M_k L_{kj} L_{ji}
\eea
as the third Casimir invariant.
In the realization of $\cal S$ by the dynamical quantities (1)-(3) of the
KP$_5$ the Casimirs are not functionally independent. Rather, they obey
the relations
\be \C_2\equiv k^2\qquad\qquad \C_3\equiv 0\qquad\qquad \C_4\equiv\C_2^2=k^4
    \qquad\qquad \C_4^{(0)}\equiv 0 \label{rel}
\ee
which, at the same time, completely fix their values. The algebra $\cal S$,
together with the relations (\ref{rel}), will be referred to as the
symmetry algebra of the classical KP$_5$.

Besides the symmetry algebra there is a well known dynamical or
non-invariance algebra (cf. \cite{BR 73} and refs.\ therein), spanned
by the functions $H_{ij}=L_{ij}$ and
\be H_{i6}= \frac12\:q_i\,\p^2 - (\q\cdot\p)\,p_i \qquad\quad
    H_{i7}=q_i \qquad\quad H_{i8}=qp_i \label{so625k1}\ee
\be H_{68}= \frac12\:q\,\p^2 \qquad\qquad\quad H_{78}=q
    \quad\qquad\qquad H_{67}=\q\cdot\p. \label{so625k2}
\ee
The Poisson algebra of the functions $H_{ab}=-H_{ba}$, $1\le a,b \le 8$,
is isomorphic to the Lie algebra so(6,2), as can be read off the Poisson
brackets of the formal linear combinations (formal, because the terms
have different dimensions) $G_{ij}=H_{ij}$, $G_{i6}=H_{i6}-\frac12\,H_{i7}$,
$G_{i7}=H_{i6} +\frac12\,H_{i7}$, $G_{i8}=H_{i8}$, $G_{67}=H_{67}$, $G_{68}=
H_{68}-\frac12\,H_{78}$ and $G_{78}=H_{68}+\frac12\,H_{78}$:
\be \big\{G_{ab},G_{ac}\big\} = g_{aa} G_{bc} \qquad\qquad
    g_{ab}=\mbox{diag}(++++++--). \label{so625k3}
\ee

\section{The Hurwitz transformation}

The Hurwitz transformation (HT) (\cite{LK 88} and refs.\ therein)
is a surjective map
\[h:\; \RP^8\longrightarrow\RP^5\qquad\qquad\quad u\longmapsto \q.\]
We will alternatively view $\RP^8=\R^8\setminus\{ 0\}$ as the space
$\HP^2$ of nonvanishing bi-quaternions and represent quaternions
$S=s_1+i\,s_2+j\,s_3+k\,s_4$ as $2\times 2$ complex matrices by putting
$1\rightarrow E_2$, $i\rightarrow -i \sigma_1$, $j\rightarrow -i\sigma_2$,
$k \rightarrow -i\sigma_3$,
where $E_2$ is the $2\times 2$ unit matrix and $\sigma_i$ are the Pauli
matrices. The multiplication of quaternions is then realized by matrix
multiplication, the conjugation $S\rightarrow \bar{S}$ by hermitian
conjugation, the real part of $S$ is Re$\,S$=$\frac12\, (S+\bar{S})$ and
the imaginary part Im$\,S$=$\frac12\, (S-\bar{S})$.

Upon the identification $\R^8=\H^2$, the components of a vector
$u=(u_1,\ldots ,u_8)$ are connected to the components of a bi-quaternion
$U=(U_1,U_2)$ via
\be U_1=\left(\begin{array}{lr} u_1-iu_4&-u_3-iu_2\\u_3-iu_2&u_1+iu_4
              \end{array}\right) \qquad
    U_2=\left(\begin{array}{lr} u_5-iu_8&-u_7-iu_6\\u_7-iu_6&u_5+iu_8
              \end{array}\right).
\ee
The Euclidean scalar product and norm on $\R^8$ appear as
\be u^\prime \cdot u=\langle U^\prime |U\rangle=\frac12 \;\mbox{tr}\,
    (U_1^\prime U_1^\dagger + U_2^\prime U_2^\dagger)
\ee
and
\be u^2:=u\cdot u=U^2:=\langle U|U \rangle=\det U_1 + \det U_2.
\ee
In terms of bi-quaternions the HT explicitly reads
\be {\sf Q}_1:=\left(\begin{array}{lr} q_1-iq_4&-q_3-iq_2\\q_3-iq_2&q_1+iq_4
              \end{array}\right) =2\, U_2U_1 \qquad
    {\sf Q}_2:=q_5 \; E_2=U_1^\dagger U_1 - U_2U_2^\dagger.
\ee
As can be seen from the identity $q=\sqrt{\q^2}=u^2$,
the HT maps spheres of radius $R$ in $\RP^8$ onto spheres of radius
$R^2$ in $\RP^5$. Geometrically the HT represents a realization of the
extension of the Hopf bundle SU(2)$\longrightarrow S^7 \longrightarrow
S^4$ to the bundle SU(2)$\longrightarrow \RP^8 \longrightarrow \RP^5$.
The fibers are 3--spheres.

The fact that the space of unit quaternions is isomorphic to the group
SU(2), $\HP = \R^+ \times$ SU(2), can be used to introduce
Euler coordinates in $\HP^2$, in terms of which the HT becomes more
transparent (cf. \cite{Iw 90}). Let
\be U_1=|u| \,\cos\frac{\chi}2 \;\hat{U}(\varphi_1,\vartheta_1,\psi_1) \qquad
    \quad
    U_2=|u| \,\sin\frac{\chi}2 \;\hat{U}(\varphi_2,\vartheta_2,\psi_2)
    \label{ek8} \ee
\be \hat{U}(\varphi,\vartheta,\psi)=\left(\begin{array}{rr}
    \displaystyle{\cos \frac\vartheta2 \, e^{
    \textstyle -i\frac{\varphi+\psi}2}} &
    \displaystyle{-i\, \sin\frac\vartheta2 \, e^{
    \textstyle -i \frac{\varphi - \psi}2}}\\
    \displaystyle{-i\, \sin\frac\vartheta2 \, e^{
    \textstyle i \frac{\varphi - \psi}2}} &
    \displaystyle{\cos \frac\vartheta2 \, e^{
    \textstyle i\frac{\varphi+\psi}2}}
    \end{array}\right) \ee
\[ 0\le\chi < \pi,\quad 0\le\vartheta_i < \pi ,\quad 0\le \varphi_i < 2\pi,
   \quad -2\pi\le \psi_i < 2\pi  \]
then
\be {\sf Q}_1 = q \,\sin\chi\:\hat{U}(\varphi,\vartheta,\psi) \qquad\qquad
    {\sf Q}_2 = q \, \cos\chi\;E_2 \label{ek5}
\ee
($q = u^2$) where $\varphi,\vartheta,\psi$ are expressed in terms of
$\varphi_1,\vartheta_1,\psi_1$ and $\varphi_2,\vartheta_2,\psi_2$ by the
addition theorem for Euler angles (see e.g. \cite{Vi 68}).

Finally, the HT can be extended to a nonbijective ``canonical" transformation
\[ \bar{h}:\; T^*\RP^8 \longrightarrow T^*\RP^5\qquad\qquad (u,v)\longmapsto
   (\q,\p).\]
The explicit expression for $\p(u,v)$, identifying $T^*\RP^8$ with $\H^2\times
\HP^2$, reads
\bea {\sf P}_1 &:=& \left(\begin{array}{rr} p_1-ip_4 & -p_3-ip_2\\p_3-ip_2 &
                                            p_1+ip_4
              \end{array}\right) = \frac1{2\,U^2}\,(V_2 U_1 + U_2 V_1)\\
     {\sf P}_2 &:=& p_5 \: E_2 = \frac1{2\,U^2} \,\mbox{Re}\,(V_1^\dagger U_1 -
              U_2 V_2^\dagger).
\eea

\section{The Hurwitz-Kepler problem}

By application of the HT the KP$_5$ can be described as the reduced form
of a singular Hamiltonian system on $T^*\RP^8$, with SU(2) acting as a
gauge group. In the sequel the singular system will be called the
Hurwitz-Kepler problem (HKP).

As a starting point we will take the Lagrangian ${\cal L}(u,\dot{u})$, which
can be obtained from the Lagrangian $L(\q,\dot{\q})$ of the KP$_5$
\be L(\q,\dot{\q})=\frac12 \:\dot{\q}^2 + \frac{k}{q} \ee
by expressing $\q$ and $\dot{\q}$ as functions of $u$ and $\dot{u}$ by means
of the HT: $\q=\q(u)$, $\dot{\q}(u,\dot{u})=(\q(u))\dot{}$. ${\cal L}
(u,\dot{u})$
reads explicitly
\be {\cal L}(u,\dot{u})=\frac12\: S_{ab}(u)\, \dot{u}_a\dot{u}_b +
    \frac{k}{u^2},\qquad\quad 1\le a,b \le 8
\ee
where
\be S_{ab}(u)=4\: \big(u^2 \,\delta_{ab} - \sum_{l=1}^3 w_a^{(l)}w_b^{(l)}\big)
\ee
\be \begin{array}{lcrrrrrrrrrl}
    w^{(1)} &=&\big(&-u_2,& u_1,&-u_4,& u_3,& u_6,&-u_5,&-u_8,& u_7&\big)\\
    w^{(2)} &=&\big(&-u_3,& u_4,& u_1,&-u_2,& u_7,& u_8,&-u_5,&-u_6&\big)\\
    w^{(3)} &=&\big(&-u_4,&-u_3,& u_2,& u_1,& u_8,&-u_7,& u_6,&-u_5&\big).
    \end{array}
\ee
The matrix $S_{ab}$ being singular
\be \det \Big| \frac{\pa^2 {\cal L}}{\pa \dot{u}_a \pa \dot{u}_b}\Big|=
    \det|S_{ab}(u)|=0 ,\qquad\quad \mbox{rank} \: S_{ab}=5,
\ee
only five of the canonical momenta
\be v_a=\frac{\pa{\cal L}}{\pa\dot{u}_a}=S_{ab}(u)\,\dot{u}_b
\ee
are independent functions of the velocities $\dot{u}_a$ and there are three
primary constraints $K_i$ \cite{Di 64}. Observing that $v\cdot w^{(i)}=0$,
the constraints can be identified as
\be K_i=\frac12\: w^{(i)}\cdot v \approx 0.
\ee
Their Poisson brackets with respect to the canonical symplectic form $\omega_8
=du \wedge dv$ generate an su(2) algebra
\be \big\{ K_i,K_j \big\}=\varepsilon_{ijk}K_k.
\ee
Following Dirac \cite{Di 64}, we have to pass to the total Hamiltonian
$H_T=v_a \dot{u}_a - {\cal L}(u,\dot{u}) + \mu_i K_i$. Making use of the
arbitrariness of the Lagrange multipliers $\mu_i$, $H_T$ can be brought
into the form
\be H_T=\frac1{8\,u^2}\: v^2 - \frac{k}{u^2} + \lambda_i K_i =:H+\lambda_i K_i
\ee
where $\lambda_i$ are still arbitrary functions. The Poisson brackets of
$H_T$ with the constraints vanish weakly, so that there are no secondary
constraints and the $K_i$ are first class \cite{Di 64}.

In terms of the quaternionic coordinates the action of the gauge group SU(2),
generated by the constraints $K_i$, can easily be
integrated. Let $g=\exp(-i\vec{\alpha}\cdot\vec{\sigma}/2)$ $\in$ SU(2)
($\vec{\sigma}$ is the vector of Pauli matrices), then
\be (U_1,U_2) \longmapsto (g\, U_1,U_2 \,g^\dagger)\qquad\qquad
    (V_1,V_2) \longmapsto (g\, V_1,V_2 \,g^\dagger). \label{su2}
\ee
Furthermore, the quaternionic coordinates allow for a complete description
of the algebra of observables (i.e. strongly gauge invariant functions).
As can be seen from (\ref{su2}), the bilinear combinations
\be U_1^\dagger U_1,\, V_1^\dagger V_1,\, U_2 U_2^\dagger,\, V_2 V_2^\dagger,\,
    V_1^\dagger U_1,\, V_2 U_2^\dagger,\, U_2 U_1,\, V_2 V_1,\, U_2 V_1,\,
    V_2 U_1
\ee
and their conjugates are invariant under the action of the gauge group SU(2).
All observables must be functions of the 28 algebraically independent
components of these elements, which can be seen to form a realization
of the Lie algebra so(6,2), analogous to that given in
(\ref{so625k1})-(\ref{so625k3}), by defining
\bea     \left(\begin{array}{rr} H_{67}-iH_{14}&-H_{13}-iH_{12}\\
         H_{13}-iH_{12}&H_{67}+iH_{14} \end{array}\right)
     &=& \frac12 \: (U_1^\dagger V_1 + V_2 U_2^\dagger)\\
         \left(\begin{array}{rr} H_{58}-iH_{23}&-H_{42}-iH_{34}\\
         H_{42}-iH_{34}&H_{58}+iH_{23} \end{array}\right)
     &=& \frac12 \: (V_1^\dagger U_1 - U_2 V_2^\dagger)\\
         \left(\begin{array}{rr} H_{15}-iH_{45}&-H_{35}-iH_{25}\\
         H_{35}-iH_{25}&H_{15}+iH_{45}\end{array}\right)
     &=& \frac12 \: (U_2 V_1 - V_2 U_1)\\
         \left(\begin{array}{rr} H_{16}-iH_{46}&-H_{36}-iH_{26}\\
         H_{36}-iH_{26}&H_{16}+iH_{46}\end{array}\right)
     &=& -\frac14 \: V_2 V_1 \\
         H_{56}\,E_2
     &=& -\frac18 \:(V_1^\dagger V_1 - V_2 V_2^\dagger) \\
         \left(\begin{array}{rr} H_{17}-iH_{47}&-H_{37}-iH_{27}\\
         H_{37}-iH_{27}&H_{17}+iH_{47}\end{array}\right)
     &=& 2\, U_2 U_1\\
         H_{57}\,E_2
     &=& U_1^\dagger U_1 - U_2 U_2^\dagger \qquad\\
         \left(\begin{array}{rr} H_{18}-iH_{48}&-H_{38}-iH_{28}\\
         H_{38}-iH_{28}&H_{18}+iH_{48}\end{array}\right)
     &=& \frac12 \:(U_2 V_1 + V_2 U_1)\\
         H_{68}\,E_2
     &=& \frac18 \:(V_1^\dagger V_1 + V_2 V_2^\dagger) \qquad\\
         H_{78}\,E_2
     &=& U_1^\dagger U_1 + U_2 U_2^\dagger . \qquad
\eea
As this so(6,2) algebra allows to generate the algebra of observables,
we shall take it as a dynamical algebra for the HKP.

\subsection{The symmetry algebra of the Hurwitz-Kepler problem}

The canonical Hamiltonian $H$ possesses 64 integrals of motion
\be F_{ab}=u_a v_b - u_b v_a \qquad\qquad A_{ab}=\frac12\: v_a v_b -
    4\,Hu_a u_b
\ee
which, together with $H$, generate an algebra $\cal A$ characterized
by the commutation relations
\bea \big\{ F_{ab},F_{ac}\big\} &=& F_{bc}\\
     \big\{ F_{ab},A_{cd}\big\} &=& \delta_{ac} A_{bd}+\delta_{ad} A_{bc}-
                                    \delta_{bc} A_{ad}-\delta_{bd} A_{ac}\\
     \big\{ A_{ab},A_{cd}\big\} &=& -2\,H\:(
                                    \delta_{ac} F_{bd}+\delta_{ad} F_{bc}+
                                    \delta_{bc} F_{ad}+\delta_{bd} F_{ac}).
\eea
The observable part of $\cal A$, i.e.\ the commutant of the constraint
algebra su(2) $\subset$ $\cal A$ in $\cal A$, is spanned by $H$ and the
manifestly SU(2)--invariant functions $L_{ij}=H_{ij}$, $1\le i,j \le 5$,
as defined above, and\footnote{The expressions for $H$, $M_i$ and $H_{ab}$
are weakly equal to the pull-backs of the corresponding quantities in the
KP$_5$ by the transformation $\bar{h}$ and will be denoted by the same
symbols.}
\bea \left(\begin{array}{rr} M_1-iM_4&-M_3-iM_2\\M_3-iM_2&M_1+iM_4
     \end{array}\right) &=&
     -\frac14 \:V_2 V_1 + 2\,H\,U_2 U_1\\
     M_5\,E_2 &=&
     -\frac18 \:(V_1^\dagger V_1 - V_2 V_2^\dagger) + H \:(
     U_1^\dagger U_1 - U_2 U_2^\dagger) \qquad
\eea
and is isomorphic to the algebra $\cal S$.

The relations (\ref{rel}), which characterize the symmetry algebra of
the KP$_5$, are now changed into relations between the Casimirs of the
algebra $\cal S$ and the Casimir of the constraint algebra
\be \C_2\equiv k^2 - 4\,H\KQ \qquad\qquad\quad \C_3 \equiv 48\,k\KQ
    \label{rel81} \ee
\be \C_4\equiv \C_2^2 + 8\,H\KQ\C_2 + 24 \,H^2 (\KQ)^2 \qquad\qquad
    \C_4^{(0)} \equiv 2\,\KQ\C_2. \label{rel82}
\ee
The symmetry algebra of the classical HKP is the algebra $\cal S$
together with the relations (\ref{rel81}) and (\ref{rel82}). For $\KQ=0$
we regain the symmetry algebra of the KP$_5$.

The above relations express the searched for functional dependencies
between the the Casimir of the constraint algebra and the Casimirs of
the true symmetry algebra $\cal S$, thereby also confirming the observable
status of $\KQ$.

\section{Quantization of the KP$_5$}

As we want to compare the result of the quantum mechanical reduction of
the HKP to the quantum theory of the KP$_5$, we shall first carry out the
quantization of the latter. This will be done in the Schr\"odinger
representation on the Hilbert space ${\cal H}_5 = {\rm L}^2(\R^5,d^5q)$.
The operators for position and momentum are
\be Q_i=q_i \quad\qquad P_i=-i\hbar\, \pa_i = -i\hbar\:\frac{\pa}{\pa q_i}
    \quad\qquad \big[ Q_i,P_j \big] = i\hbar\,\delta_{ij}.
\ee
For the operators $H$ and $L_{ij}$ there are
no factor ordering problems, the factor ordering for the $M_i$ can be fixed
by requiring that the commutator algebra of $H$, $L_{ij}$ and $M_i$ be
isomorphic to the classical Poisson algebra. The so-obtained expressions
\be H=\frac12 \:\P^2 - \frac{k}{Q} \quad\qquad L_{ij}=Q_i P_j - Q_j P_i \ee
\be M_i=Q_i \,(\frac12 \:\P^2 + H) - (\Q\cdot\P)\,P_i + 2i\hbar P_i
\ee
($Q=\sqrt{\Q^2}=\sqrt{\q^2}$) with the commutation relations
\be \big[H,L_{ij}\big] = 0 = \big[H,M_i\big] \qquad\qquad
\big[L_{ij},L_{ik}\big] = i\hbar L_{jk} \ee
\be \big[L_{ij},M_i\big]=i\hbar M_j \qquad\qquad
    \big[M_i,M_j\big]=i\hbar (-2H)L_{ij}
\ee
are simultaneously hermitian with respect to the measure $d^5q$.
The classical identities (\ref{rel}) for the Casimirs of the algebra
$\cal S$ acquire quantum corrections of order $\hbar^2$
\bea \C_2 &=& \M^2 - 2H L^2 \equiv k^2 + 8 \,\hbar^2 H \qquad\quad
     \C_3  = 6\,\varepsilon_{ijklm} L_{ij}L_{kl}M_m \equiv 0
             \qquad \label{rel5q1} \\
     \C_4 &=& 2H^2 L_{ij}L_{jk}L_{kl}L_{li} +
              \frac12 \,(M_iM_iM_jM_j + M_iM_jM_jM_i) + \nonumber\\
          & & + H\,(M_i L_{ij} L_{jk} M_k +
              M_i M_j L_{jk} L_{ki} +
              L_{ij} M_j M_k L_{ki} + L_{ij} L_{jk} M_k M_i) \nonumber\\
          &\equiv& \C_2^2 - 12 \,\hbar^2 H \C_2\\
     \C_4^{(0)} &=& \M^2 L^2 - \frac14 (M_i L_{ij}+L_{ij} M_i)(M_k L_{kj}+
                    L_{kj} M_k) \equiv -4\,\hbar^2 \C_2. \label{rel5q2}
\eea
Consequently, the symmetry algebra of the quantized KP$_5$ is the algebra
$\cal S$ together with these modified relations.
Note that the terms in the defining expression for $\C_4$ do not reflect
factor ordering ambiguities but are uniquely determined by the ordering
of the terms according to the original classical expression
$\C_4 = \frac12 \,l_{\mu\nu}l_{\nu\rho}l_{\rho\sigma}l_{\sigma\mu}$.

As the symmetry algebra does not contain all the dynamical information,
we shall in addition require that the dynamical algebra generated by $H_{ab}$
be represented on ${\cal H}_5$ as a commutator algebra. By this requirement
the (hermitian) expressions for the operators $H_{ab}$ are found to
be\footnote{The five-dimensional generalizations of the operators $L_{ab}$
of ref. \cite{BR 73}, which are hermitian with respect to the measure
$q^{-1}d^5q$, can be obtained from our $H_{ab}$ (resp.\ $G_{ab}$, which are
formed from $H_{ab}$ as above (\ref{so625k3})) by replacing $P_i$ with $\Pi_i
= -i\hbar\,(\pa_i - \frac{q_i}{2q^2})$. The operators $\Pi_i$ are canonically
conjugate to the $Q_j$ and hermitian with respect to $q^{-1}d^5q$.}
\bea H_{i6} &=& \frac12\:Q_i \P^2 - (\Q\cdot\P)\,P_i + i\hbar\:\frac52\,P_i
                - \hbar^2\:\frac18\,\frac{Q_i}{Q^2} \qquad\qquad
                H_{i7}=Q_i \quad \label{1}\\
     H_{i8} &=& Q P_i - i\hbar\:\frac12\,\frac{Q_i}{Q} \qquad\qquad\qquad
                H_{67}=\Q\cdot\P - i\hbar\:\frac52 \label{2}\\
     H_{68} &=& \frac12\,Q\,\P^2 - i\hbar\:\frac12\,\frac1{Q}\:\Q\cdot\P -
                \hbar^2\:\frac78\,\frac1{Q} \qquad\qquad\qquad H_{78}=Q.
                \label{3}
\eea
Note that a mere symmetrization of the classical expressions would not
have resulted in the desired closing algebra.

In order to determine the energy spectrum and a basis of ${\cal H}_5$
which is well adapted to the \rep s of the symmetry algebra, we shall
separate the Schr\"odinger equation in the Euler coordinates defined by
(\ref{ek5}). In these coordinates the wave functions are the simultaneous
eigenfunctions of the complete set of commuting observables $H$, $L^2$,
$J^2=2\,(\JQ_1+\JQ_2)$, $J_{1_3}$ and $J_{2_3}$, where the angular
momentum vectors
\be \vec{J}_1=\frac12 \left(\begin{array}{c} L_{12}+L_{34}\\L_{13}+L_{42}\\
              L_{14}+L_{23} \end{array}\right) \quad\qquad
    \vec{J}_2=\frac12 \left(\begin{array}{c} L_{12}-L_{34}\\L_{13}-L_{42}\\
              L_{14}-L_{23} \end{array}\right) \label{so4} \ee
generate an SO(4) subgroup of SO(5). The angular wave functions
$Y_{jm_1m_2}^l$ $(\chi,\varphi,\vartheta,\psi)$ are labelled according to the
group chain SO(5) $\supset$ SO(4) $\cong$ SU(2)$\times$SU(2)
$\supset$ U(1)$\times$U(1), and satisfy the eigenvalue equations
\bea J_{1_3} \: Y_{jm_1m_2}^l
     &=& -i\hbar \,\pa_\varphi \:\Yjmml = -\hbar m_1 \,\Yjmml\\
     J_{2_3} \: \Yjmml
     &=& -i\hbar \,\pa_\psi \:\Yjmml = -\hbar m_2\, \Yjmml\\
     J^2 \: \Yjmml
     &=& -4\hbar^2(\pa_\vartheta^2 + \cot \vartheta \, \pa_\vartheta +
         \frac1{\sin^2 \vartheta}\,(\pa^2_\varphi + \pa^2_\psi -
         2 \cos \vartheta \, \pa_\varphi \pa_\psi))\, \Yjmml \nonumber\\
     &=& 4\hbar^2 \, j(j+1)\, \Yjmml\\
     L^2 \:\Yjmml
     &=& - \hbar^2\,(\pa^2_\chi + 3 \cot \chi \,\pa_\chi -
         \frac1{\sin^2 \chi} \frac1{\hbar^2}\, J^2)\, \Yjmml \nonumber\\
     &=& \hbar^2 \, l(l+3)\, \Yjmml.
\eea
They are explicitly given by
\be \Yjmml(\chi,\varphi,\vartheta,\psi)=N_j^l\, \sin^{2j}\chi \;
    C_{l-2j}^{2j+\frac32}(\cos\chi)\: D^j_{m_1m_2}(\varphi,\vartheta,\psi)
\label{Yl}\ee
where $N_j^l$ is a normalization constant, $C_{n}^{\alpha}(z)$
are Gegenbauer polynomials \cite{AS 65} and
\be D^j_{m_1m_2}(\varphi,\vartheta,\psi)=e^{-im_1\varphi} \, P^j_{m_1m_2}
    (\cos\vartheta)\, e^{-im_2\psi} \label{wf}
\ee
is the ($m_1,m_2$)--matrix element in the representation $D^j$ of the
element $g$ $\in$ SU(2), which is described by the Euler angles
($\varphi,\vartheta,\psi$) (cf. \cite{Vi 68}).

The energy spectrum and the radial wave functions $R_{Nl}(q)$ are determined
by the radial Schr\"odinger equation
\be \Big( -\frac{\hbar^2}2 \Big(\pa_q^2 + \frac4{q}\,\pa_q - \frac{l(l+3)}{q^2}
    \Big)-\frac{k}q \Big)\: R_{Nl}(q)=E_N \,R_{Nl}(q). \label{rs5}
\ee
The regular (as $q \rightarrow 0$) solutions of this equation can be obtained
in the same way as in the three-dimensional case (cf. \cite{LL 62}). For
$E<0$, = 0 and $>0$ they are proportional to Laguerre polynomials, Bessel
functions and confluent hypergeometric functions respectively (see
\cite{AS 65})
\bea R_{nl}(q) &=& N_{nl}\: e^{-\frac12 \rho} \: \rho^l \:
                   L^{2l+3}_{n-l-1}(\rho)  \qquad\qquad\qquad
                   \rho = \frac{2k}{\hbar^2(n+1)}\: q\\
     R_{0l}(q) &=& N_{0l}\: \rho^{-\frac32}\: J_{2l+3} (\sqrt{\rho})
                   \qquad\qquad\qquad\qquad\qquad\,
                   \rho = \frac{8k}{\hbar^2}\:q\\
  R_{\nu l}(q) &=& N_{\nu l}\: e^{-\frac12 \rho} \: \rho^l \:
                      _1\!F_1(l+2+i\nu,2l+4;\rho)  \qquad
                   \rho = i \frac{2k}{\hbar^2 \nu} \: q.
\eea
The $N_{Nl}$ are normalization constants. Of course, for $E\ge 0$, the wave
functions are improper eigenstates of $H$ and can only be normalized to
delta functions. The energy spectrum is of the form
\be E_n=-\frac{k^2}{2\hbar^2\,(n+1)^2} \qquad\qquad E_0=0 \qquad\qquad
    E_\nu=\frac{k^2}{2\hbar^2\nu^2}.
\ee
The range of the quantum numbers $N=(n,0,\nu),\, l,\, j,\, m_1$ and $m_2$,
needed to uniquely label the states, is as follows
\be 0 < n \in {\bf N},\qquad\qquad 0 < \nu \in \R \ee
\be 0\le l \le n-1 \quad (E<0),\qquad\qquad 0\le l \quad (E \ge 0) \ee
\be 0\le 2j \le l, \qquad\qquad -j\le m_1,m_2 \le j. \ee
$l$ is integer, $m_1$ and $m_2$ are, simultaneously with $j$, both integers
or both half integers. With respect to the measure $d^5q$ the states are
orthogonal in all quantum numbers.

\subsection{Group theoretical considerations \label{gtc}}

The Hamiltonian $H$ of the KP$_5$ possesses an obvious symmetry under the
canonical action of the
group SO(5) of rotations in $\R^5$. This means, that its eigenvalues
cannot depend on the quantum numbers $j$, $m_1$ and $m_2$, which label
the states within an irreducible representation of SO(5). But the energy
eigenvalues are also degenerate in the angular momentum quantum number
$l$ and the eigenspaces of $H$ are not irreducible with respect to SO(5).
This ``accidental" degeneracy is due to the higher symmetry of the KP$_5$,
which reflects itself in the existence of the additional conserved vector
$\M$, and can be explained by an invariance of the Hamiltonian
under the groups SO(6), E(5) and SO(5,1) for $H<0$, $H=0$ or $H>0$
respectively\footnote{E(5) is the group of motions in five-dimensional
Euclidean space, also called the inhomogeneous rotation group ISO(5).}
(cf. \cite{BI 66}).
Consequently, the eigenspaces of $H$ carry irreducible representations of
the three groups. However, not all the \ir\ \rep s of the three groups do
occur as energy eigenspaces. The physically realized \rep s are selected
by the relations between the Casimir operators of the symmetry algebra, which
at the same time fix their eigenvalues and allow to express the spectrum of
$H$ by the group quantum numbers. In the following we shall determine the
relevant unitary \ir\ \rep s (UIR) of the above groups, which simultaneously
furnish the \ir\ \rep s of the symmetry algebra of the KP$_5$.

As $H$ is central in $\cal S$, it must be represented by a multiple of
unity in the \ir\ \rep s of $\cal S$. In the eigenspaces of $H$, i.e.
for $H=E=$ const., the algebra $\cal S$ can be reduced to a trivial central
extension of the Lie algebras so(6), e(5) and so(5,1) by replacing $M_i$ with
\be \tilde{M}_i = L_{i6}=-L_{6i} = \left\{\begin{array}{rclcl}
    (-2H)^{-\frac12} \: M_i & \qquad & E&<&0\\
    M_i & & E&=&0.\\
    (2H)^{-\frac12} \: M_i & & E&>&0
    \end{array}\right. \label{so6}
\ee
The commutation relations of the operators $L_{\mu \nu}$ ($1\le \mu,\nu\le 6$)
read
\be \big[ L_{\mu\nu},L_{\mu\rho}\big]=i\hbar\,g_{\mu\mu}L_{\nu\rho} \ee
\be g_{\mu\nu}=\mbox{diag}\:(1,1,1,1,1,\varepsilon), \qquad
    \varepsilon = -\mbox{sign}\, (E). \ee
Upon this reduction the identities (\ref{rel5q1})-(\ref{rel5q2}), which
hold for the Casimirs of $\cal S$, induce relations for the Casimirs of
the three Lie algebras. Therefore, the hermitian \ir\ \rep s of the
symmetry algebra of the KP$_5$ are uniquely determined by those \ir\ \rep s
of the Lie algebras so(6), e(5) and so(5,1), in which these induced
relations are satisfied, and which correspond to UIR of the groups SO(6),
E(5) and SO(5,1).

The three cases will be treated separately. For the UIR of the groups
SO(6), E(5) and SO(5,1) and the eigenvalues of their Casimir operators
consult the appendix.
\begin{itemize}
\item $E<0$

For negative energies we have the relations
\bea C_2 &=& \frac1{2\hbar^2}\: L_{\mu\nu}L_{\mu\nu}=\frac1{\hbar^2}\:
             (L^2 + \tilde{\M}^2) \equiv
             -\frac1{2\hbar^2 H}\:(k^2 + 8\hbar^2 H)\\
     C_3 &=& \frac1{\hbar^3}\: \varepsilon_{\mu\nu\rho\sigma\tau\upsilon}
             L_{\mu\nu}L_{\rho\sigma}L_{\tau\upsilon}=\frac6{\hbar^3}\,
             \varepsilon_{ijklm} L_{ij}L_{kl}\tilde{M}_m \equiv 0\\
     C_4 &=& \frac1{2\hbar^4}\:
L_{\mu\nu}L_{\nu\rho}L_{\rho\sigma}L_{\sigma\mu}
             \equiv C_2^2 + 6\: C_2.
\eea
The last two identities require that the SO(6) quantum numbers $\mu_2$ and
$\mu_3$ (cf. (A 15)-(A 17)) be zero, the first one allows to express the energy
by the eigenvalue of the Casimir $C_2$
\be H=E=-\frac{k^2}{2\hbar^2(C_2+4)}=-\frac{k^2}{2\hbar^2(\mu_1+2)^2}.
\ee
This means that the energy quantum number $n$ is connected to the SO(6) quantum
number $\mu_1$ via $n=\mu_1+1$. The $H$--eigenspace belonging to energy $E_n$
carries the \rep\ D($n-1,0,0$) of SO(6) and the branching rules for SO(6)
$\supset$ SO(5)
\be \mu_1 \ge l \ge 0 \quad \Longleftrightarrow \quad n-1 \ge l \ge 0
\ee
(see (A 3), $m_{5,1}=\mu_1$, $m_{5,2}=\mu_2=0$, $m_{4,1}=l$) correctly account
for the ``accidental" degeneracy.

\item $E=0$

For $E=0$ the relations
\bea C_2 &=& \frac1{k^2}\: L_{i6}L_{i6}=\frac1{k^2}\: \tilde{\M}^2 \equiv 1 \\
     C_3 &=& \frac1{k\hbar^2}\: \varepsilon_{\mu\nu\rho\sigma\tau\upsilon}
             L_{\mu\nu}L_{\rho\sigma}L_{\tau\upsilon}=\frac6{k\hbar^2}\:
             \varepsilon_{ijklm} L_{ij}L_{kl}\tilde{M}_m \equiv 0 \\
C_4^{(0)}&=& \frac1{k^2\hbar^2}\: (L^2\tilde{\M}^2 - \frac14 (\tilde{M}_i
             L_{ij} + L_{ij} \tilde{M}_i)(\tilde{M}_k L_{kj} + L_{kj}
             \tilde{M}_k) \equiv -4\: C_2 \qquad
\eea
can only be satisfied by putting $\sigma =1$, $\mu_2=\mu_3=0$ (cf.
(A 21)-(A 23)). The 0--eigenspace of $H$ carries the \rep\ D(1;0,0) of E(5),
and the accidental degeneracy is explained by the branching rules for
E(5) $\supset$ SO(5):
\be l \ge 0 \ee
($m_{5,1}=\infty$, $m_{5,2}=\mu_2=0$, $m_{4,1}=l$ in (A 3)).

\item $E>0$

For $E>0$ we have ($\tilde{L}_{ij}=L_{ij}$, $\tilde{L}_{i6}=-iL_{i6}=-i
\tilde{M}_i$)
\bea C_2 &=& \frac1{2\hbar^2}\: \tilde{L}_{\mu\nu}\tilde{L}_{\mu\nu}=
             \frac1{\hbar^2}\: (L^2 - \tilde{\M}^2) \equiv
             -\frac1{2\hbar^2 H}\:(k^2 + 8\hbar^2 H) \label{*}\\
     C_3 &=& \frac{i}{\hbar^3}\: \varepsilon_{\mu\nu\rho\sigma\tau\upsilon}
             \tilde{L}_{\mu\nu}\tilde{L}_{\rho\sigma}\tilde{L}_{\tau\upsilon}=
             \frac6{\hbar^3}\,\varepsilon_{ijklm} L_{ij}L_{kl}\tilde{M}_m
             \equiv 0 \\
     C_4 &=& \frac1{2\hbar^4}\: \tilde{L}_{\mu\nu}\tilde{L}_{\nu\rho}
             \tilde{L}_{\rho\sigma}\tilde{L}_{\sigma\mu}
             \equiv C_2^2 + 6\: C_2. \label{**}
\eea
The requirement $H>0$, together with the identity (\ref{*}), restricts us
to the \rep s of the principal series. Within the principal series the
relations (\ref{*})-(\ref{**}) are only compatible with $\mu_1=\mu_2=-1$,
$\mu_3=i\tau$, $\tau > 0$ (cf. (A 15)-(A 17) and the remark below). Solving
equation (\ref{*}) for $H$, we obtain
\be H=E=-\frac{k^2}{2\hbar^2 (C_2 +4)} = \frac{k^2}{2\hbar^2 \tau^2}.
\ee
The energy quantum number $\nu$ is equal to the SO(5,1) quantum number $\tau$
and the eigenspace of $H$ corresponding to energy $E_\nu$ carries the \rep\
D($p;-1,-1,i\nu$) of SO(5,1). Again, the ``accidental" degeneracy is correctly
reproduced by the branching rules for SO(5,1) $\supset$ SO(5), which require
that $l\ge 0$.
\end{itemize}

As the range of the quantum numbers $j$, $m_1$ and $m_2$ is in accordance
with the branching rules for the subgroup chain SO(5) $\supset$ SO(4)
$\cong$ SU(2)$\times$SU(2) $\supset$ U(1)$\times$U(1) in the \rep\ D($l$,0)
of SO(5) (cf. the appendix), we see that the quantum numbers $N$, $l$, $j$,
$m_1$ and $m_2$ uniquely label the states within the above UIR of the groups
SO(6), E(5) and SO(5,1).

\section{Quantization of the Hurwitz-Kepler problem}

In this section we will perform the quantization of the HKP and its
reduction according to the algebraic constraint quantization scheme.
The first step is the quantization of the extended system, without
imposing the constraints, and the determination of the \ir\ \rep s of
the symmetry algebra which span the Hilbert space. The second step consists
in the identification of the physical \rep s and of the physical Hilbert
space. The resulting quantum theory will be compared to that of the
KP$_5$.

\subsection{Hilbert space and observables}

The quantization of the HKP will be performed in the Schr\"odinger \rep\
\be u_a \longrightarrow u_a \qquad\qquad v_a \longrightarrow -i\hbar\,\pa_a
    =-i\hbar\:\frac{\pa}{\pa u_a}
\ee
on the Hilbert space ${\cal H}_8={\rm L}^2(\R^8,d\mu(u))$, which
represents the appropriate kinematical setting corresponding to the
symplectic structure with respect to which the classical HKP is
defined. The measure $d\mu(u)$ will be determined later on by physical
requirements.

As the fundamental subalgebras of the algebra of observables to be
quantized we shall choose the symmetry algebra $\cal S$ and the dynamical
algebra generated by $H_{ab}$. As the latter linearly encodes
the structure of the algebra of observables, we shall require that the
corresponding commutator algebras be isomorphic to the classical Poisson
algebras. For the symmetry algebra this can be achieved by simply putting
all $v$'s to the right of all $u$'s in the classical expressions and
replacing them with operators, for the $H_{ab}$ we obtain the more
complicated expressions
\bea     \left(\begin{array}{rr} H_{16}-iH_{46}&-H_{36}-iH_{26}\\
         H_{36}-iH_{26}&H_{16}+iH_{46}\end{array}\right)
     &=& -\frac14 \,(-i\hbar)^2 D_2 D_1 + i\hbar\,\frac{(-i\hbar)}{4\,U^2}\:
         (U_2 D_1 + D_2 U_1) \nonumber\\
     & & - \hbar^2 \frac1{4\,U^4}\:U_2 U_1\\
         \left(\begin{array}{rr} H_{17}-iH_{47}&-H_{37}-iH_{27}\\
         H_{37}-iH_{27}&H_{17}+iH_{47}\end{array}\right)
     &=& 2\, U_2 U_1\\
         \left(\begin{array}{rr} H_{18}-iH_{48}&-H_{38}-iH_{28}\\
         H_{38}-iH_{28}&H_{18}+iH_{48}\end{array}\right)
     &=& \frac12\,(-i\hbar) \,(U_2 D_1 + D_2 U_1) - i\hbar\,\frac1{U^2}\,
         U_2 U_1
\eea
\bea     H_{56}\,E_2
     &=& -\frac18\,(-i\hbar)^2\,(D_1^\dagger D_1 - D_2 D_2^\dagger) +
         i\hbar \,\frac{(-i\hbar)}{4\,U^2}\,\mbox{Re}\,(U_1^\dagger D_1
         - U_2 D_2^\dagger) \nonumber\\
     & & - \hbar^2 \frac1{8\,U^4}\:(U_1^\dagger U_1 - U_2 U_2^\dagger)\\
         H_{57}\,E_2
     &=& U_1^\dagger U_1 - U_2 U_2^\dagger \qquad\\
         H_{58}\,E_2
     &=& \frac12\,(-i\hbar)\,\mbox{Re}\,(U_1^\dagger D_1 - U_2 D_2^\dagger)
         - i\hbar\,\frac1{2\,U^2}\,(U_1^\dagger U_1 - U_2 U_2^\dagger)\\
         H_{67}\,E_2
     &=& \frac12\,(-i\hbar)\,\mbox{Re}\,(U_1^\dagger D_1 + U_2 D_2^\dagger)
         - i\hbar\:\frac52 \:E_2\\
         H_{68}\,E_2
     &=& \frac18\,(-i\hbar)^2 \,(D_1^\dagger D_1 + D_2 D_2^\dagger)
         - i\hbar\,\frac{(-i\hbar)}{4\,U^2}\,\mbox{Re}\,(U_1^\dagger D_1 + U_2
         D_2^\dagger) \nonumber\\
     & & - \hbar^2\: \frac78\,\frac1{U^2} \:E_2\\
         H_{78}\,E_2
     &=& U_1^\dagger U_1 + U_2 U_2^\dagger
\eea
where
\be D_1=\left(\begin{array}{lr} \pa_1-i\pa_4&-\pa_3-i\pa_2\\
        \pa_3-i\pa_2&\pa_1+i\pa_4 \end{array}\right) \qquad
    D_2=\left(\begin{array}{lr} \pa_5-i\pa_8&-\pa_7-i\pa_6\\
        \pa_7-i\pa_6&\pa_5+i\pa_8 \end{array}\right).
\ee
Although it is possible to consistently quantize the constraint algebra
in the same way as the symmetry algebra, we will not make use of it
except for comparative purposes.

In order to find the operator corresponding to the observable $\KQ$, the
Casimir of the constraint algebra, we start from the classical identity
$\C_3=48\,k\KQ$. For the theory to possess the correct classical limit,
in leading order in $\hbar$ this identity must be reproduced. The possible
correction terms must contain explicit powers of $\hbar$ and be polynomials
in $H$, $\C_2$, $\KQ$ and $k$. The only polynomial combination of these
elements with the correct dimensions is the term $\hbar^2 k$. As can easily
be seen from the defining expression for $\C_3$, such a term cannot occur
in it. Therefore we conclude that the relation between $\C_3$ and $\KQ$
remains unaltered. Using the so-obtained expression for $\KQ$, which
coincides with the one that can be inferred from the quantization of the
constraint algebra, the quantum analogs of the relations (\ref{rel81})-%
(\ref{rel82}) for the Casimirs of the symmetry algebra are found to be
(the defining expressions are the same as in (\ref{rel5q1})-(\ref{rel5q2}))
\bea \C_2 &\equiv& k^2 - 4\,H\KQ + 8\,\hbar^2 H \qquad\qquad\quad
                   \C_3\equiv 48\,k\KQ \label{rel8q1} \\
     \C_4 &\equiv& \C_2^2 + 8\,H\KQ\C_2 + 24\,H^2(\KQ)^2 - 12\,\hbar^2 H
                   \C_2 - 48\,\hbar^2 H^2 \KQ \qquad\\
\C_4^{(0)}&\equiv& 2\,\KQ\C_2 - 4\,\hbar^2\C_2. \label{rel8q2}
\eea
The symmetry algebra of the quantized HKP is the algebra $\cal S$,
supplemented by these relations. For $\KQ=0$ it becomes isomorphic
to the symmetry algebra of the quantized KP$_5$.

The measure $d\mu(u)$ can be determined by two physically motivated
requirements: $i$) it should be invariant under the action of the linear
(in $V$) part of the symmetry algebra, generated by $L_{ij}$; $ii$) the
observables should be hermitian with respect to it. The first requirement
can be satisfied by putting
\be d\mu(u)=c\,|u|^n d^8u, \qquad c=\mbox{const}.,\ee
the second can then only be fulfilled if $n=2$. Thus we are led to choose
the space ${\cal H}_8={\rm L}^2(\R^8, c\,u^2 d^8u)$ as the extended Hilbert
space for the HKP.

A basis of ${\cal H}_8$ which is well suited for the explicit construction
of the \rep s of the symmetry algebra and which facilitates the comparison
of the quantum theory of the HKP to that of the KP$_5$ can be obtained by
separating the Schr\"odinger equation in the bi-Euler coordinates
(\ref{ek8}). In these coordinates the Laplace operator has the form
\bea \triangle^{(8)} &=& \pa^2_\upsilon + \frac7\upsilon \:\pa_\upsilon +
                         \frac4{\upsilon^2}\: \big(\pa^2_\chi + 3\,\cot\chi\:
                         \pa_\chi + \frac1{\cos^2\frac\chi2}\:\triangle_
                         {\vartheta_1\varphi_1\psi_1} + \frac1{\sin^2
                         \frac\chi2}\:\triangle_{\vartheta_2\varphi_2\psi_2}
                         \big) \nonumber \\
                     &=:& \pa^2_\upsilon + \frac7\upsilon \:\pa_\upsilon -
                         \frac1{\hbar^2\upsilon^2}\: F^2 \\
     \triangle_{\vartheta\varphi\psi}
                     &=& \pa^2_\vartheta + \cot\vartheta\:\pa_\vartheta +
                         \frac1{\sin^2\vartheta} \big( \pa^2_\varphi +
                         \pa^2_\psi - 2\, \cos\vartheta\:\pa_\varphi\pa_\psi
                         \big)
\eea
($\upsilon=|u|$). Besides the energy quantum number, there are seven more
quantum numbers needed to label the states, corresponding to the group chain
SO(8) $\supset$ SO(4)$\times$SO(4) $\supset$ U(1)$\times$U(1)$\times$U(1)$
\times$U(1), where SO(4)$\times$SO(4) is generated by the angular momentum
vectors ($F_{ab}=-i\hbar\,(u_a \pa_b - u_b \pa_a)$)
\be \vec{I}_1=\frac12 \left(\begin{array}{c} F_{12}+F_{34}\\F_{13}+F_{42}\\
              F_{14}+F_{23} \end{array}\right) \qquad\qquad
    \vec{I}_2=\frac12 \left(\begin{array}{c} F_{12}-F_{34}\\F_{13}-F_{42}\\
              F_{14}-F_{23} \end{array}\right) \ee
\be \vec{N}_1=\frac12 \left(\begin{array}{c} F_{56}+F_{78}\\F_{57}+F_{86}\\
              F_{58}+F_{67} \end{array}\right) \qquad\qquad
    \vec{N}_2=\frac12 \left(\begin{array}{c} F_{56}-F_{78}\\F_{57}-F_{86}\\
              F_{58}-F_{67} \end{array}\right). \ee
The angular wave functions $\Yf(\chi,\varphi_1,\vartheta_1,\psi_1,\varphi_2,
\vartheta_2,\psi_2)$ are the simultaneous eigenfunctions of the mutually
commuting operators $F^2$, $I^2=2\,(\IQ_1+\IQ_2)$, $N^2=2\,(\NQ_1+\NQ_2)$,
$I_{1_3}$, $I_{2_3}$, $N_{1_3}$ and $N_{2_3}$ with the eigenvalues
\bea F^2\:Y^f &=& \hbar^2 f(f+6)\:Y^f \\
     I^2\:Y^f &=& -4\,\hbar^2 \triangle_{\vartheta_1\varphi_1\psi_1}\, Y^f =
                   4\,\hbar^2 j_1(j_1+1)\: Y^f \\
     N^2\:Y^f &=& -4\,\hbar^2 \triangle_{\vartheta_2\varphi_2\psi_2}\, Y^f =
                   4\,\hbar^2 j_2(j_2+1)\: Y^f \\
 I_{1_3}\:Y^f &=& -i\hbar\,\pa_{\varphi_1}\: Y^f = -\hbar m_{11}\: Y^f \\
 I_{2_3}\:Y^f &=& -i\hbar\,\pa_{\psi_1}\: Y^f = -\hbar m_{12}\: Y^f \\
 N_{1_3}\:Y^f &=& -i\hbar\,\pa_{\varphi_2}\: Y^f = -\hbar m_{21}\: Y^f \\
 N_{2_3}\:Y^f &=& -i\hbar\,\pa_{\psi_2}\: Y^f = -\hbar m_{22}\: Y^f.
\eea
Their explicit form is
\bea \Yf &=& N^f_{j_1j_2}\: (1+\cos\chi)^{j_1} (1-\cos\chi)^{j_2}\,
             P_{\frac{f}2 -j_1-j_2}^{(2j_2+1,2j_1+1)}(\cos\chi)
             \times\nonumber\\
         & & \times D^{j_1}_{m_{11}m_{12}}(\varphi_1,\vartheta_1,\psi_1)
             \:D^{j_2}_{m_{21}m_{22}}(\varphi_2,\vartheta_2,\psi_2)
\eea
where $P_n^{(\alpha,\beta)}(z)$ are Jacobi polynomials \cite{AS 65} and
$D^j_{mn}(\varphi,\vartheta,\psi)$ are Wigner functions as defined in
(\ref{wf}).

The spectrum of $H$ and the radial wave functions are determined by the radial
Schr\"odinger equation. By making the substitution $q=\upsilon^2=|u|^2$
the latter can be brought into the same form as the radial Schr\"odinger
equation for the KP$_5$ (\ref{rs5})
\be \Big( -\frac{\hbar^2}2 \Big(\pa_q^2 + \frac4{q}\,\pa_q - \frac1{q^2}
    \frac{f}2\Big(\frac{f}2 +3\Big)
    \Big)-\frac{k}q \Big)\: R_{N\frac{f}2}(q)=E_N \,R_{N\frac{f}2}(q)
    \label{rs8} \ee
with $l$ replaced by $\frac{f}2$. Correspondingly, the spectrum and the
radial wave functions coincide with those of the KP$_5$, but for negative
energies the quantum number $n$ can now take on also half integer values.
Simultaneously with $2n$, $f$ has to be even or odd. The quantum numbers
$N=(n,0,\nu)$, $f$, $j_1$, $j_2$, $m_{11}$, $m_{12}$, $m_{21}$ and $m_{22}$
obey the inequalities
\be 2\le 2n \in {\bf N}, \qquad\quad 0<\nu \in \R \ee
\be \left.\begin{array}{lcl} 0\le f\le 2n-2 & & \mbox{$n$ integer,
    $f$ even} \\ 1\le f \le 2n-1 & & \mbox{$n$ half integer, $f$ odd}
    \end{array}\right\} (H<0), \qquad 0\le f \in {\bf N} \quad (H\ge 0)
\ee
\be \left.\begin{array}{lcl} 0\le j_1+j_2 \le\frac{f}2 & &
    \mbox{$f$ even, $j_1+j_2$ integer}\\\frac12\le j_1+j_2 \le\frac{f}2 & &
    \mbox{$f$ odd, $j_1+j_2$ half integer} \end{array}\right\} \quad
    -j_i \le m_{i1},m_{i2} \le j_i. \ee
With respect to the measure $u^2d^8u$ the wave functions are orthogonal
in all quantum numbers.

\subsection{Representations of the symmetry algebra}

Just as in the case of the KP$_5$, in the eigenspaces of $H$ the symmetry
algebra of the HKP can be reduced to a trivial central extension of the Lie
algebras so(6), e(5) or so(5,1), and its \ir\ \rep s can be inferred from
the UIR of the groups SO(6), E(5) and SO(5,1). The relevant \rep s, from
which the Hilbert space ${\cal H}_8$ is made up, and the spectrum of $\KQ$
are determined by the relations (\ref{rel8q1})-(\ref{rel8q2}) between the
Casimirs of the symmetry algebra and the Casimir of the constraint algebra.

In order to obtain the spectrum of $\KQ$, we start from the identities for
the Casimirs of the Lie algebra e(5) which are induced by the relations
(\ref{rel8q1})-(\ref{rel8q2}) (the definitions of the generators and the
Casimirs of the three Lie algebras are the same as in the case of the KP$_5$)
\be C_2 \equiv 1 \qquad\qquad C_3 \equiv \frac{48}{\hbar^2}\:\KQ \qquad\qquad
    C_4^{(0)} \equiv \frac2{\hbar^2}\:\KQ C_2 - 4\,C_2.
\ee
The first relation requires that $\sigma = 1$, the second and third that
$\mu_2 = \mu_3 =: K$ (cf. (A 21)-(A 23)). Therefore, from the second
relation, the spectrum of $\KQ$ is of the form $\hbar^2\,K(K+1)$,
$2 K \in {\bf N}$, in accordance with the spectrum of the Casimir of
the constraint algebra su(2). The zero-eigenspace ${\cal H}_0$ of $H$
decomposes into a direct sum of UIR of E(5) according to
\be {\cal H}_0 = \bigoplus_{2\,K=0}^\infty {\rm D}(1;K,K).
\ee
Similarly, for negative and positive eigenvalues of $H$ the eigenspaces
${\cal H}_N$ decompose into direct sums of UIR of the groups SO(6) and
SO(5,1), and the spectrum can be expressed by the group quantum numbers.
We shall indicate below the induced identities for the Casimirs of the Lie
algebras so(6) and so(5,1), the form of the energy spectrum, and the
\rep s which are contained in the corresponding energy eigenspaces.

\begin{itemize}
\item $E<0: \qquad\quad (\KQS=\frac1{\hbar^2}\:\KQ)$
\bea C_2 &\equiv& -\frac1{2\,\hbar^2 H}(k^2 - 4\,H\KQ + 8\,\hbar^2 H) \\
     C_3 &\equiv& \frac{48\,k}{\hbar^3\sqrt{-2\,H}}\:\KQ \\
     C_4 &\equiv& C_2^2 + 6\,C_2 - 4\,\KQS C_2 - 12\,\KQS + 6\,(\KQS)^2 \\
       H &=& E=-\frac{k^2}{2\,\hbar^2(C_2+4-2\,\KQS)}=-\frac{k^2}{2\,\hbar^2
             (\mu_1+2)^2}\\
{\cal H}_n &=& \bigoplus_{K=0}^{n-1} {\rm D}(n-1,K,K) \qquad\quad
                    \mbox{$\mu_1=n-1$ integer}\\
{\cal H}_n &=& \bigoplus_{K=\frac12}^{n-1} {\rm D}(n-1,K,K) \qquad\quad
                    \mbox{$\mu_1=n-1$ half integer}\quad
\eea
\item $E>0$
\bea C_2 &\equiv& -\frac1{2\,\hbar^2 H}(k^2 - 4\,H\KQ + 8\,\hbar^2 H) \\
     C_3 &\equiv& \frac{48\,k}{\hbar^3\sqrt{2\,H}}\:\KQ \\
     C_4 &\equiv& C_2^2 + 6\,C_2 - 4\,\KQS C_2 - 12\,\KQS + 6\,(\KQS)^2 \\
       H &=& E=-\frac{k^2}{2\,\hbar^2 (C_2+4-2\,\KQS)} = \frac{k^2}{2\,
             \hbar^2\tau^2} \\
{\cal H}_\nu &=& \bigoplus_{2\,K=0}^\infty {\rm D}(p;-1+K,-1+K,\nu)
                   \qquad\quad \tau = \nu ,\,\tau>0. \quad
\eea
\end{itemize}

As can be seen from the above decomposition of the eigenspaces of $H$, the
Hilbert space ${\cal H}_8$ is the direct sum (integral) of all those UIR
of the groups SO(6), E(5) and SO(5,1), which are compatible with the relations
(\ref{rel8q1})-(\ref{rel8q2}), i.e.\ of all hermitian \ir\ \rep s of the
symmetry algebra of the HKP. The basis states for the corresponding \rep\
spaces, which diagonalize the mutually commuting operators $H$, $L^2$, $J^2=
2\,(\JQ_1+\JQ_2)$, $\tilde{J}^2=8\,(\JQ_1-\JQ_2)$, $J_{1_3}$, $J_{2_3}$,
$\KQ$ and $K_3$ (the vectors $\vec{J}_i$ are defined as in (\ref{so4})),
can easily be constructed from the basis of ${\cal H}_8$ as given in the
previous paragraph.

First of all, observe that the vectors $\vec{J}_i$ and $\vec{K}$ can be
expressed as linear combinations of the vectors $\vec{I}_i$ and $\vec{N}_i$
\be \vec{J}_1=\vec{N}_1, \qquad\quad \vec{J}_2=\vec{I}_2, \qquad\quad
    \vec{K}=\vec{I}_1 - \vec{N}_2.
\ee
Therefore, and because of the identities $\IQ_1=\IQ_2$, $\NQ_1=\NQ_2$, the
angular wave functions $\Yf$ are already eigenfunctions of the operators $J^2$,
$\tilde{J}^2$, $J_{1_3}$, $J_{2_3}$ and $K_3$ with the eigenvalues
$2\,\hbar^2(j_1(j_1+1) + j_2(j_2+1))$, $8\,\hbar^2(j_1(j_1+1) - j_2(j_2+1))$,
$-\hbar\:m_{21}$, $-\hbar\:m_{12}$ and $-\hbar\:(m_{11}-m_{22})$. To get
the eigenfunctions of $\KQ$, we simply have to apply the rules for the
addition of angular momenta. They can be written as the Clebsch--Gordan
series
\be \YK := \sum_{m_{11}=-j_1}^{j_1}\sum_{m_{22}=-j_2}^{j_2}
           \langle j_1m_{11}j_2(-m_{22})|KM \rangle\:
           Y_{j_1m_{11}m_2j_2m_1m_{22}}^{2l} \label{cgs}
\ee
($f=2l$, $m_1=m_{21}$, $m_2=m_{12}$, $M=m_{11}-m_{22}$,
$|j_1-j_2|\le K \le j_1+j_2$) and satisfy the eigenvalue equations
\bea \KQ\:\YK &=& \hbar^2K(K+1)\:\YK \\
     K_3\:\YK &=& -\hbar \,M\:\YK.
\eea
Because of the identity
\be F^2 \equiv 4\:(L^2-\KQ), \ee
these states are also eigenstates of $L^2$ with eigenvalue
$\hbar^2(l(l+3) + K(K+1))$.
Thus, the desired basis states are $R_{Nl}(q)\, \YK(\chi,\varphi_1,
\vartheta_1,\psi_1,\varphi_2,\vartheta_2,\psi_2)$.

\subsection{Algebraic reduction and physical Hilbert space}

According to the algebraic constraint quantization scheme outlined in
the introduction, the physical Hilbert space $\Hp$ is the direct sum of
(the \rep\ spaces of) those UIR of the symmetry algebra of the HKP, in
which the Casimir of the constraint algebra has the value
zero. This definition is consistent because $\KQ$ is a Casimir
of the algebra of observables, and can therefore be represented by a
multiple of unity, and because zero is contained in its spectrum.
For $\KQ=0$ the relations (\ref{rel8q1})-(\ref{rel8q2}) between the
Casimirs of the algebra $\cal S$ and the Casimir of the constraint
algebra induce relations which must be satisfied by the Casimirs of
$\cal S$. Consequently, the physical Hilbert space is spanned by those \ir\
\rep s of $\cal S$, in which these relations hold.

Using the material of the previous section, the physical \rep s can
easily be found. For $\KQ=0$ we have $K=0$, and we are left with the
following \rep s (identifying the \rep s with their carrier spaces)
\be \begin{array}{lclcll}
    {\cal H}_{n,phys} &=& {\rm D}(n-1,0,0) &\qquad& E_n=-\displaystyle{
    \frac{k^2}{2\,\hbar^2(n+1)^2}} \quad & \mbox{$n$ integer}\\
    {\cal H}_{0,phys} &=& {\rm D}(1;0,0) & & E_0=0 & \\
    {\cal H}_{\nu,phys} &=& {\rm D}(p;-1,-1,i\nu) & &
    E_\nu=\displaystyle{\frac{k^2}{2\,\hbar^2\nu^2}}&
    \end{array}
\ee
which coincide with the \rep s of the symmetry algebra of the KP$_5$ as
given in section \ref{gtc}.

The physical Hilbert space is spanned by the basis states of the above
\rep s, i.e. by the states $R_{Nl}(q)\: Y_{jm_1jm_200}^l$ ($K=M=0$,
$j_1=j_2=:j$). Clearly, these states also constitute a basis of the
space of SU(2)-invariant states. Furthermore, $\Hp$ can isomorphically
be mapped onto the Hilbert space L$^2(\R^5,d^5q)$ of the KP$_5$, as can be
seen as follows.

First observe that for $j_1=j_2=j$ the Jacobi polynomials are proportional
to Gegenbauer polynomials \cite{AS 65} and the $\chi$-dependent functions
become
\bea (1+\cos\chi)^j (1-\cos\chi)^j\,P_{l - 2j}^{(2j+1,2j+1)}(\cos\chi) &=&
     \frac{(4j+2)!(l+1)!}{(2j+1)!(2j+l+2)!}\:
     \sin^{2j}\chi \times\nonumber\\
     & & \times \, C^{2j+\frac32}_{l-2j}(\cos\chi).
\eea
Then, from the properties of Clebsch--Gordan coefficients,
$\langle jm_{11}j(-m_{22}) | 00 \rangle \sim \delta_{m_{11}m_{22}}$,
and the group \rep\ property of Wigner functions
\be \sum_{m=-j}^j D_{m_1m}^j(\varphi_1,\vartheta_1,\psi_1)\:D_{mm_2}^j
    (\varphi_2,\vartheta_2,\psi_2) = D_{m_1m_2}^j(\varphi,\vartheta,\psi)
\ee
($\varphi,\vartheta,\psi$ being connected to $\varphi_1,\vartheta_1,
\psi_1,\varphi_2,\vartheta_2,\psi_2$ via the addition theorem
for Euler angles) we see that the Clebsch-Gordan series (\ref{cgs}) yields
\be Y_{jm_1jm_200}^l = {\cal N}_j^l\: \sin^{2j}\chi\;C^{2j+\frac32}_{l-2j}
    (\cos\chi)\;D_{m_1m_2}^j (\varphi,\vartheta,\psi).
\ee
Thus, the gauge invariant states depend only on the gauge invariant
variables $q_i$ (\ref{ek5}), and equations (\ref{Yl}) and (\ref{rs8})
show that they coincide with the eigenfuctions of the KP$_5$ in Euler
coordinates. Therefore, the bases of the spaces $\Hp$ and L$^2(\R^5,d^5q)$
can be mapped onto one another in a one-to-one manner.

Furthermore, the induced measure on $\Hp$ can be shown to be proportional
to $d^5q$. For this purpose we use the 1-forms
\be dq_i=\frac{\pa q_i}{\pa u_a}\:du_a, \qquad\qquad
    \kappa_i = 2\,w_a^{(i)}\,du_a.
\ee
The 1-forms $\kappa_i$ are ``dual" to the constraint vector fields
\be X_i=\frac12 \: w_a^{(i)} \frac\pa{\pa u_a}, \qquad\qquad
    \kappa_i(X_j)=u^2\delta_{ij}.
\ee
In terms of $dq_i$ and $\kappa_i$ the measure $c\,u^2d^8u$ can be
written as
\be c\,u^2d^8u = \frac{c}{2^8} \: d^5q\wedge\kappa_1\wedge\kappa_2
    \wedge\kappa_3,
\ee
where $\kappa_1\wedge\kappa_2\wedge\kappa_3$ is the volume form on the
gauge group SU(2). The induced measure on $\Hp$ is obtained by integrating
over the gauge group, and, with $c=\frac{2^7}{\pi^2}$, is equal to $d^5q$,
showing that $\Hp$ and L$^2(\R^5,d^5q)$ are isomorphic as Hilbert spaces.
The integration over the group can also be interpreted as the projector
onto the gauge invariant states.

\subsection{Reduction of observables}

In order to fully establish the equivalence of the quantum theories of
the HKP and of the KP$_5$, we still have to prove that the \rep s of the
fundamental observables on $\Hp$ and on L$^2(\R^5,d^5q)$ are equivalent.
This will be done by showing that the generators of the symmetry algebra
and of the dynamical algebra act in the same way on the basis states of
the two spaces. For the generators $H$, $L_{ij}$ and $M_i$ of the symmetry
algebra this is clear from the group theoretical treatment. For the
generators of the dynamical algebra it is proved if we can show it for
the operators $H_{a8}$, because all other elements of the algebra can be
obtained from them by means of the commutation relations. The operators
$H_{a8}$ can be expressed in terms of the operators $\KQ$, $Q_i=Q_i(u)=q_i(u)$
and $P_i$, where
\bea \left(\begin{array}{rr} P_1-iP_4 & -P_3-iP_2\\P_3-iP_2 & P_1+iP_4
     \end{array}\right) &=& \frac{(-i\hbar)}{2\,U^2}\,(U_2 D_1 + D_2 U_1)\\
     P_5 \: E_2 &=& \frac{(-i\hbar)}{2\,U^2} \,\mbox{Re}\,(U_1^\dagger D_1 -
              U_2 D_2^\dagger),
\eea
as
\bea H_{i8} &=& Q P_i - i\hbar \,\frac12\,\frac{Q_i}{Q} \qquad\qquad\qquad\,
                H_{78}=Q=q\\
     H_{68} &=& \frac12\,Q\,\P^2 + \frac1{2\,Q}\,\KQ - i\hbar \,\frac12
                \frac1{Q}\, \Q\cdot\P - \hbar^2\frac78\frac1{Q}.
\eea
Therefore, because of (\ref{2}) and (\ref{3}), it suffices to show that the
action of $\Q$ and $\P$ on $\Hp$ coincides with that on L$^2(\R^5,d^5q)$.
For the operators $Q_i$ this follows immediately from (\ref{ek5}), because
they have the same form on both spaces in Euler coordinates. For the $P_i$
it can be seen from the commutation relations of $\Q$ and $\P$,
$\big[ Q_i,P_j \big]=i\hbar\,\delta_{ij}$, and the fact that the Hamiltonian
can be written as
\be H = \frac12\:\P^2 - \frac{k}{Q} + \frac1{2\,Q^2}\:\KQ,
\ee
which allows to represent $P_i$ on $\Hp$ as
\be P_i=\frac{i}{\hbar}\:\big[ H,Q_i \big]
\ee
and to infer the action of $P_i$ from that of $H$ and $Q_i$.

\section{Conclusions}

As we have demonstrated, the algebraic approach to the quantization of
constrained systems provides a powerful and elegant tool for the
quantization and reduction of the HKP. The interpretation of (the
vanishing of) the Casimir of the constraint algebra as a \rep\
condition on the physical \rep s of the symmetry algebra could very
effectively be used for the construction of the physical Hilbert space.
The quantum expression for the Casimir of the constraint algebra and its
spectrum could be determined intrinsically, without having to quantize
the constraint algebra, thereby also confirming its interpretation as an
observable.

To summarize, it may be said that, by laying emphasis on observable
quantities, the algebraic method is closer to the physical interpretation
of gauge systems than is the more kinematical method of implementing the
individual constraints as projectors onto the physical states, and that
the difficulties connected with the \rep\ of the unphysical constraint
algebra (Dirac's bit of luck \cite{Di 64}) can be completely avoided.

\newpage
\begin{appendix}
\setcounter{equation}{0}
\renewcommand{\theequation}{\mbox{A \arabic{equation}}}

\section*{Appendix: The UIR of SO(6), SO(5,1) and E(5)}

In this appendix we give a brief characterization of the unitary irreducible
representations (UIR) of the groups G = SO(6), SO(5,1), E(5) (=ISO(5)) by
means of the eigenvalues of their Casimir invariants and the branching rules
for the restriction of the representations according to the subgroup
chain G $\supset$ SO(5) $\supset$ SO(4) $\supset$ SO(3) $\supset$ SO(2).
For the details see the cited literature.

\subsection*{The UIR of SO(6)}

The UIR of SO(6) can be characterized by the components of the corresponding
highest weight $(m_{5,1},m_{5,2},m_{5,3})$, where the $m_{5,i}$ are all
integers or all half integers which satisfy the inequality
\be m_{5,1} \ge m_{5,2} \ge |m_{5,3}|. \ee
The representations will be denoted D($m_{5,1},m_{5,2},m_{5,3}$). The
branching rules for the subgroup chain SO(6) $\supset$ SO(5) $\supset$
SO(4) $\supset$ SO(3) $\supset$ SO(2) can be read off the Gelfand-Zetlin
pattern \cite{BR 80}
\be \left|\begin{array}{ccccc} m_{5,1}& &m_{5,2}& &m_{5,3}\\ &m_{4,1}& &
m_{4,2}& \\ &m_{3,1}& &m_{3,2}& \\ & &m_{2,1}& & \\ & &m_{1,1}& &
\end{array}\right| ,\ee
where $m_{i,j}$ denotes the j-th component of the highest weights for the
representations of SO(i+1) which are contained in the representation
D($m_{5,1},m_{5,2},m_{5,3}$) upon subsequent restriction to these subgroups.
The $m_{i,j}$ are all integers or all half integers and satisfy the
inequalities
\be \begin{array}{l} m_{5,1} \ge m_{4,1} \ge m_{5,2} \ge m_{4,2} \ge
|m_{5,3}|\\
m_{4,1} \ge m_{3,1} \ge m_{4,2} \ge m_{3,2} \ge -m_{4,2} \\
m_{3,1} \ge m_{2,1} \ge |m_{3,2}| \\
m_{2,1} \ge m_{1,1} \ge -m_{2,1}.  \end{array}\ee
All representations, whose highest weights are compatible with these
inequalities, occur exactly once.

\subsection*{The UIR of SO(5,1)}

The UIR of SO(5,1) can also be labelled by a Gelfand-Zetlin pattern as in
(A 2), but now the range of the numbers $m_{5,i}$ and the branching rules
for SO(5,1) $\supset$ SO(5) are different. The branching rules for SO(5)
$\supset$ SO(4) $\supset$ SO(3) $\supset$ SO(2) are the same as in the case of
SO(6). There are three series of UIR (see \cite{Ot 68}, the numbers $l_{i,j}$
of Ottoson are connected with the $m_{i,j}$ by $l_{2r,k}=m_{2r,k}+r-k+1$,
$l_{2r-1,k}=m_{2r-1,k}+r-k$):
\begin{enumerate}
\item The principal series: D(p; $m_{5,1},m_{5,2},i \tau$)
\be m_{5,1} \ge m_{5,2} \ge -1 ,\qquad m_{5,3}=i\tau, \qquad \tau \in
{\bf R}\ee
\be m_{4,1} \ge m_{5,1}+1 \ge m_{4,2} \ge m_{5,2}+1 .\ee
$m_{5,i}$ and $m_{4,i}$ are all integers or all half integers. If $m_{5,2}
=-1$, then $\tau > 0$.
\item The supplementary series: D(s; $m_{5,1},-1,1$)
\be m_{5,1} \ge -1 , \qquad m_{5,2}=-m_{5,3}=-1 \ee
\be m_{4,1} \ge m_{5,1}+1 ,\qquad m_{4,2}=0. \ee
Here $m_{5,1}$ and $m_{4,1}$ are both integers.
\item The exceptional series: D(e; $m_{5,1},-1,m_{5,3}$)
\be m_{5,1} \ge -1,\qquad m_{5,2}=-1,\qquad 0<m_{5,3}<1  \ee
\be m_{4,1} \ge m_{5,1}+1 \ge m_{4,2} \ge 0.\ee
Here $m_{5,1}$ and $m_{4,i}$ are all integers.
\end{enumerate}
Again, each of the representations of the subgroups, whose highest weights
are compatible with the branching rules, occurs exactly once.

\subsection*{The UIR of E(5)}

The Gelfand-Zetlin pattern for the UIR of E(5) can be obtained from
that of SO(6) by making $m_{5,1} \longrightarrow \infty$. In addition,
one needs a continuous parameter $\sigma \in {\bf R}$ (see \cite{Ch 68}).
The range of the numbers $m_{i,j}$, the branching rules, and the
multiplicities of the representations of the subgroups are the same
as for SO(6). The representations will be denoted D($\sigma$; $m_{5,2},
m_{5,3}$).

\subsection*{The Casimir operators and their eigenvalues}

The generators of the Lie algebras so(6), so(5,1) and e(5) and their
commutation relations can be written in a uniform manner. Let
\be L_{\mu\nu} = - L_{\nu\mu},\qquad 1\le\mu,\nu\le 6 \ee
then
\be \big[ L_{\mu\nu},L_{\mu\rho} \big] = i g_{\mu\mu} L_{\nu\rho} \ee
where $g_{ii}=1$ for $1\le i \le 5$,  $g_{66}=1$, -1, 0 for so(6), so(5,1),
e(5) resp., and $g_{\mu\nu} = 0$ for $\mu\neq\nu$. Each of the algebras
possesses three Casimir operators. For so(6) they can be
chosen as
\bea C_2 &=& \frac12\, L_{\mu\nu} L_{\mu\nu}\\
     C_3 &=& \varepsilon_{\mu\nu\rho\sigma\tau\upsilon} L_{\mu\nu}
             L_{\rho\sigma} L_{\tau\upsilon}\\
     C_4 &=& \frac12 \,L_{\mu\nu} L_{\nu\rho} L_{\rho\sigma} L_{\sigma\mu} \eea
with the eigenvalues (see \cite{PP 66, Ok 77})
\bea C_2 &=& \mu_1(\mu_1 +4) + \mu_2(\mu_2+2) + \mu_3^2\\
     C_3 &=& 48 \: (\mu_1+2)(\mu_2+1)\mu_3 \label{i}\\
     C_4 &=& (\mu_1(\mu_1+4))^2 + 6 \mu_1(\mu_1+4) + (\mu_2(\mu_2+2))^2 +
             \mu_3^4 - 2 \mu_3^2
\qquad \eea
where $\mu_i=m_{5,i}$ as given above. The Casimirs for so(5,1) can be obtained
from those of so(6) by making the substitution $L_{i6}\rightarrow -i\,
L_{i6}$ and multiplying the defining expression for $C_3$ by $+i$. The
eigenvalues in terms of $\mu_1$, $\mu_2$, $\mu_3$ are the same, apart from a
factor $-i$ on the right hand side of (\ref{i}).

For e(5) the Casimir invariants are \cite{WY 79}
\bea C_2 &=& L_{i6} L_{i6}\\
     C_3 &=& \varepsilon_{\mu\nu\rho\sigma\tau\upsilon} L_{\mu\nu}
             L_{\rho\sigma} L_{\tau\upsilon}\\
     C_4 &=& \frac12 \,L_{ij}L_{ji}L_{6k}L_{k6} - \frac14 \,(L_{i6}L_{ij} +
             L_{ij}L_{i6})(L_{k6}L_{kj} + L_{kj}L_{k6}). \qquad\eea
Their eigenvalues are
\bea C_2 &=& \sigma^2\\
     C_3 &=& 48 \,\sigma (\mu_2+1)\mu_3\\
     C_4 &=& \sigma^2 (\mu_2(\mu_2+2) + \mu_3^2 - 4).\eea
The UIR of SO(6) and E(5) are uniquely determined by the eigenvalues of
their Casimir operators, in the case of SO(5,1) the additional knowledge of
the series or the branching rules for SO(5,1) $\supset$ SO(5) is necessary.

Finally, we want to give the Casimir invariants for the subalgebras
so(5) and so(4) and their eigenvalues. For so(5), generated by $L_{ij}$,
$1\le i,j \le 5$, there are two Casimirs
\bea L^2 &=& \frac12 \,L_{ij}L_{ij} \\
C_4^{(5)}&=& \frac12 \,L_{ij}L_{jk}L_{kl}L_{li} \eea
with the eigenvalues ($\lambda_1=m_{4,1}$, $\lambda_2=m_{4,2}$)
\bea L^2 &=& \lambda_1(\lambda_1 + 3) + \lambda_2(\lambda_2 + 1) \\
C_4^{(5)}&=& (\lambda_1(\lambda_1+3))^2 + 3\lambda_1(\lambda_1+3) +
             (\lambda_2(\lambda_2+1))^2- \lambda_2(\lambda_2+1). \eea
The UIR of SO(5), which are uniquely specified by the numbers $\lambda_1$
and $\lambda_2$, will be denoted D($\lambda_1,\lambda_2$).

For so(4) = su(2)$\oplus$su(2), generated by the two commuting angular
momentum vectors
\be \vec{J}_1=\frac12 \left(\begin{array}{c} L_{12}+L_{34}\\L_{13}+L_{42}\\
              L_{14}+L_{23} \end{array}\right)\qquad\qquad
    \vec{J}_2=\frac12 \left(\begin{array}{c} L_{12}-L_{34}\\L_{13}-L_{42}\\
              L_{14}-L_{23} \end{array}\right) \ee
the Casimir invariants are ($1\le i,j,k,l \le 4$)
\bea J^2 &=& \frac12 \,L_{ij}L_{ij} = 2\,(\vec{J}_1^2+\vec{J}_2^2)\\
 \tilde{J}^2 &=& \varepsilon_{ijkl}L_{ij}L_{kl}=8\,(\vec{J}_1^2-\vec{J}_2^2).
\eea
They have the eigenvalues ($m_{3,1}=j_1+j_2$, $m_{3,2}=j_1-j_2$)
\bea J^2 &=& 2 \,(j_1(j_1+1)+j_2(j_2+1))\\
\tilde{J}^2 &=& 8\,(j_1(j_1+1)-j_2(j_2+1)).\eea
Again, the UIR are uniquely labelled by $j_1$ and $j_2$. They will be
denoted ${\rm D}^{j_1j_2}$.

\end{appendix}

\end{document}